**Starlink Satellite Brightness –**

**Characterized From 100,000 Visible Light Magnitudes**

Anthony Mallama

anthony.mallama@gmail.com

2021 November 16

Abstract

Magnitudes for the VisorSat and Original-design types were analyzed separately and by time. Mean values are compared with those from other large-scale photometric studies, and some signficant differences are noted. The illumination phase functions for Starlink satellites indicate strong forward scattering of sunlight. They are also time-dependent on a scale of months and years. These phase functions improve the predictability of satellite magnitudes. A Starlink Brightness Function tailored to the satellite shape also improves magnitude predictions. Brightness flares lasting a few seconds are characterized and the mean rate of magnitude variation during a pass is determined. Observation planning tools, including graphs and statistics of predicted magnitudes, are discussed and illustrated.

## 1. Introduction

Observational astronomy is being impacted by the large number of bright satellites in low-earth-orbit (Hall et al., 2021, Walker et al., 2020a, Walker et al., 2020b, Tyson et al., 2020, Otarola et al., 2020, Gallozzi et al., 2020, Hainaut and Williams, 2020, McDowell, 2020, Williams et al. 2021a, Williams et al. 2021b). More than a thousand Starlink communication satellites have been launched by the SpaceX company during the past two years and many more are planned. Other companies such as OneWeb and Amazon, as well as the government of China are pursuing similar projects.

In collaboration with the astronomical community, SpaceX developed a VisorSat model of Starlink that includes a Sun shade to make the satellites fainter. This paper characterizes the observed brightness of VisorSat spacecraft and Original-design Starlink satellites in visible light. More than 60,000 VisorSat magnitudes and 40,000 Original magnitudes from the database of the MMT-9 automated observatory were specially processed and then analyzed for this study.

Characterizing Starlink satellite brightness allows for the prediction of their magnitudes. This capability can aid astronomers in planning their observations to minimize interference from bright satellites.

Section 2 describes the hardware, the magnitudes and the database of the MMT-9 automated observatory. Section 3 examines the apparent magnitudes of Starlink satellites and their dispersion. Section 4 reports on the satellite magnitudes after adjustment to a standard distance of 1,000 km. Section 5 characterizes the magnitudes by fitting them to a Phase Function (PF) and demonstrates that this equation changes with time. The PFs for VisorSats and for Original satellites reduce the uncertainties of their magnitudes.

Section 6 describes the development and use of a physical model called the Starlink Brightness Function (SBF). The section begins by defining a satellite-centered coordinate system (SCCS) which corresponds to the shape of a Starlink spacecraft. The SCCS is the frame of reference for the SBF. Then the computation of data needed for the SBF is addressed. The required data includes the positions of the Sun and the MMT-9 in the SCCS system. Next, the evaluation of the parameter coefficients for the SBF is described. Finally, the ability of the SBF to improve magnitude predictions by reducing their uncertainties is illustrated.

Section 7 discusses the prediction of satellite magnitudes and illustrates the brightness distribution of Starlink satellites across the sky. Section 8 characterizes Starlink flares which are short duration brightness surges and addresses the rate of time varying brightness changes normally seen during a satellite pass. Section 9 reports on the brightness of satellites that have failed operationally but are still in orbit and on other satellite anomalies. Section 10 lists the limitations of this study. Section 11 discusses the research reported here in the context of other similar studies. The conclusions from this analysis are summarized in Section 12.

**2. MMT-9 hardware and observations**

Mini-MegaTORTORA (MMT-9) is located in Russia at the site of their 6-m telescope (Beskin et al. 2017). This robotic observatory includes nine 71 mm diameter f/1.2 lenses and 2160 x 2560 sCMOS sensors. The detectors are sensitive to the visible spectrum from red through blue. S. Karpov (private communication) provided a color transformation formula which indicates that MMT-9 magnitudes taken through the clear filter are within 0.1 magnitude of the Johnson-Cousins V band-pass for objects with solar-like color indices. The transformation equation to convert from MMT-9 magnitudes to V-band is provided in Appendix A.

The [MMT-9 database](#) (Karpov et al. 2015) contains photometry of satellite passes collected into *track* files. Starlink files containing over 100,000 magnitude records were downloaded for this study. The observations were recorded between September 18, 2019 and September 9, 2021. They were separated by time into the 10 Groups listed in Table 1 so that monthly and yearly changes could be investigated.

Table 1. Observation Groups by date

```
VisorSat
#    Start Date    End Date      Obs.    Scatter
4    2020-Sep-21   2020-Dec-29   10,076   0.16
8    2021-Jan-01   2021-Mar-15   10,408   0.21
1    2021-Jan-03   2021-May-08   10,628   0.19
9    2021-Apr-07   2021-Jun-20   10,308   0.20
6    2021-Jun-11   2021-Sep-04   10,444   0.20
0    2021-Jul-02   2021-Sep-09   10,576   0.22

Original
#    Start Date    End Date      Obs.    Scatter
5    2019-Sep-18   2019-Nov-18   10,304   0.17
3    2020-Apr-19   2020-Dec-30   10,557   0.11
2    2021-Jan-04   2021-Mar-10   10,257   0.16
7    2021-Apr-30   2021-Aug-27   10,143   0.19
```

The MMT-9 observations were recorded when satellites were more than 20° above the horizon. The distribution of satellite azimuths relative to that of the Sun was reasonably uniform as shown in Figure 1.

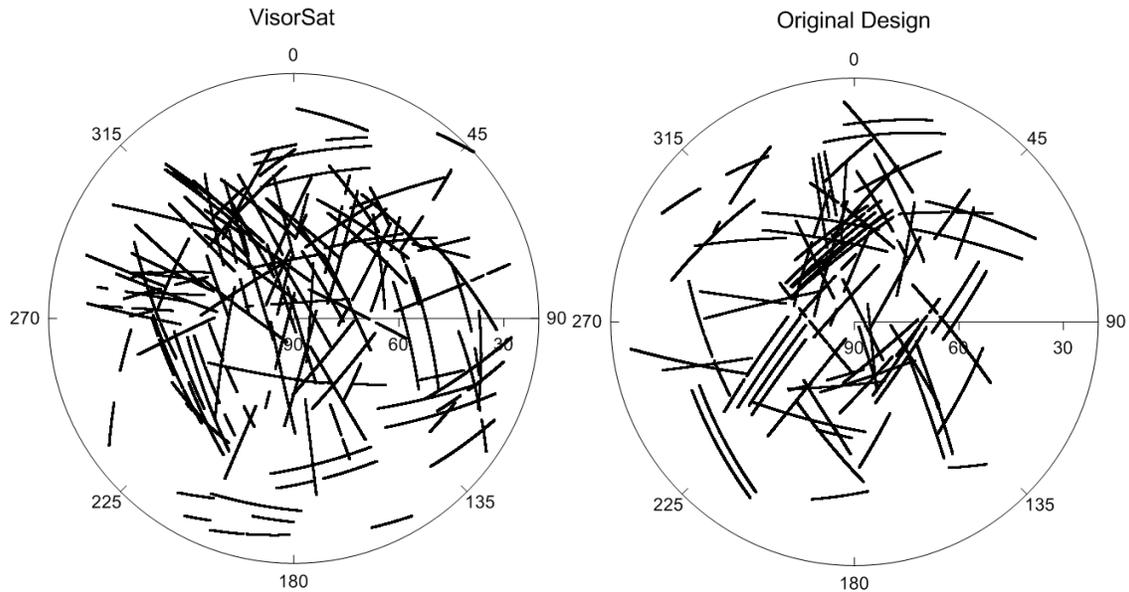

Figure 1. The distribution of satellite tracks in the sky plotted as elevation above the horizon and azimuth relative to the Sun.

The MMT-9 magnitudes include a moderate amount of scatter. There are 28,287 instances of time-adjacent observations where the same satellite was recorded by different instrument channels within one second of time. The overall RMS difference of those observation pairs is 0.18 magnitude. The scatter for each Group of observations is listed in Table 1.

All spacecraft were at their operational altitude (approximately 550 km) at the times of the observations according to the plots maintained by J. McDowell. Passes containing magnitudes recorded while the satellites were in the Earth's penumbral shadow were excluded.

## 3. Apparent magnitudes

Astronomical observations are impacted directly by the *apparent magnitudes* of satellites. These are the values of brightness as recorded by a sensor and corrected for atmospheric extinction alone. Table 2 lists mean apparent magnitudes for VisorSats and Original-design satellites by Group number in order of date. The first row below the individual Groups, labeled 'Unified', lists the results of processing all the observations in one solution. The next line, labeled 'Aggregated', gives the results of combining the individual Group solutions. The Aggregated standard deviations are less than the Unified values because the means of the Group solutions vary. Finally, on the line labeled 'No Scatter', the standard deviation from the Aggregated solution has been adjusted to account for the values of scatter listed in Table 1.

Table 2. Apparent magnitudes

```
VisorSat
#     Year      Mag     SD   SD_Mean
4    2020.90   6.34    0.39   0.00
8    2021.12   6.29    0.70   0.01
1    2021.27   6.10    0.97   0.01
9    2021.41   6.84    0.86   0.01
6    2021.52   6.56    0.98   0.01
0    2021.59   6.48    0.89   0.01
Unified        6.43    0.86   0.00
Aggregated     6.43    0.82   0.00
No Scatter             0.80   0.00

Original
#     Year      Mag     SD   SD_Mean
5    2019.79   5.02    0.46   0.00
3    2020.68   5.14    0.55   0.01
2    2021.10   4.89    0.47   0.00
7    2021.48   5.16    0.56   0.01
Unified        5.05    0.52   0.00
Aggregated     5.05    0.51   0.00
No Scatter             0.49   0.00

SD = standard deviation
SD_Mean = SD of the mean
```

The mean apparent magnitude for VisorSat observations is 6.43 and that for Originals is 5.05, so the difference is 1.38 magnitudes. The apparent magnitudes for VisorSats are also distinguished from Originals by having a greater dispersion. The No Scatter standard deviation of the former is 0.80 magnitude while that of the latter is 0.49. One notable exception for VisorSats is Group 4, the earliest set of those magnitudes. The standard deviation for Group 4 is smaller than those for all the others groups including Originals. The standard deviations of the mean magnitudes for each Group are 0.00 or 0.01 due to the large number of observations. Figure 2 illustrates the comparison between VisorSat and Original apparent magnitudes as a bar chart.

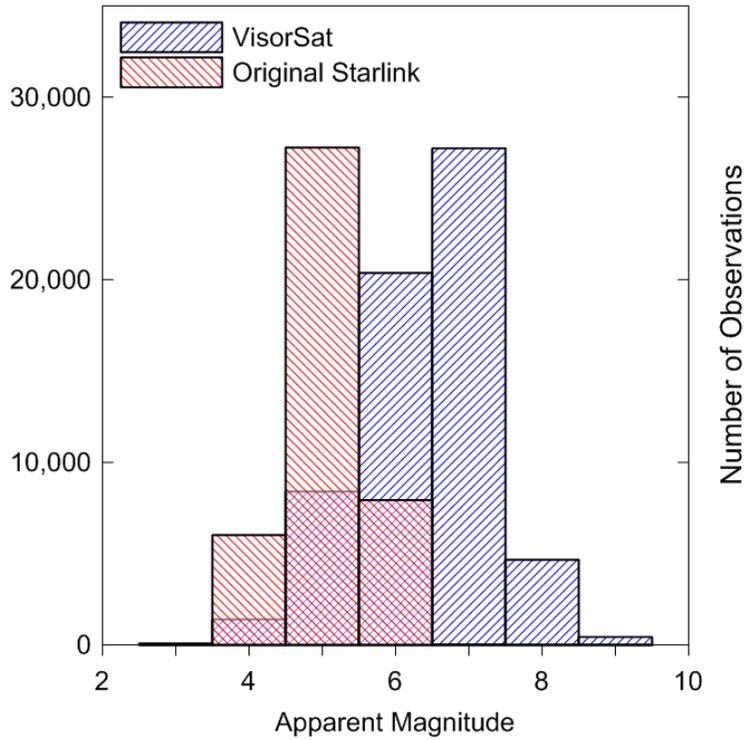

Figure 2. The apparent magnitudes of VisorSats are substantially fainter than those of the Original Starlink satellites.

The mean apparent magnitudes from each Group are plotted versus time in Figure 3. The faintest value for VisorSat, from Group 9 at year 2021.41, is addressed in the Section 5 on phase angles. The duration of observation for Original satellites is seen to be about 2 years while that of the newer VisorSats is only 1 year.

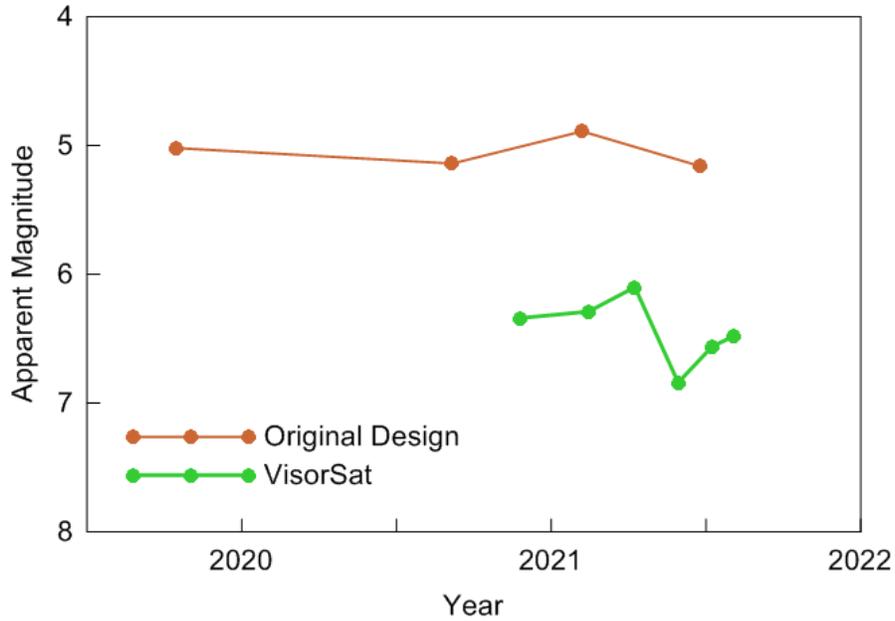

Figure 3. Apparent magnitude means of Starlink satellite Groups plotted versus the year of observation.

**4. Magnitudes at a standard distance**

An analogy to the absolute magnitude of astronomers is the *magnitude at a standard distance* used by satellite analysts. This value is calculated by applying the inverse-square law of light to the apparent magnitude. A distance of 1000 km is commonly chosen as the standard and it is convenient for comparisons between multiple satellite constellations at different altitudes. Some analysts prefer to use 550 km for Starlink satellites because that is their approximate orbital altitude. The difference between brightness values adjusted to these two distances is 1.30 magnitudes. Table 3 lists mean 1000-km magnitudes for VisorSats and Original satellites by Group. The lines labeled Unified, Aggregated and No Scatter are explained in Section 3.

The mean 1000-km magnitude for VisorSats is 7.21 and that for Original satellites is 5.89. The difference of 1.32 magnitudes implies that the VisorSats average 30% as bright as the Original satellites. This compares to the value of 31% reported by Mallama (2021).

VisorSats may also be compared in brightness to OneWeb satellites. Mallama (2020) computed a mean 1000-km magnitude of 7.18 +/-0.03 for OneWeb spacecraft based on MMT-9 observations. So, the difference is only 0.03 magnitude. When the 1000-km value is adjusted to the nominal 1,200 km altitude of a OneWeb satellite in orbit, it corresponds to magnitude 7.58. Meanwhile, VisorSats at their 550 km nominal altitude average magnitude 5.91. Thus, VisorSat satellites are approximately the same brightness as OneWeb at a common distance but they are much brighter than OneWeb at their respective operational altitudes.

Adjusting the magnitudes to a standard distance might be expected to reduce the large dispersions of the apparent magnitudes. However, the No Scatter standard deviation for VisorSats actually increases to 0.81 (from 0.80) and that for the Originals only reduces to 0.43 (from 0.49). An explanation for this unexpected result is offered in the next section. As with apparent magnitudes, Group 4 is a notable exception as its standard deviation is much smaller than those for the other groups of VisorSats. The standard deviations for the mean magnitudes of each Group are 0.00 or 0.01 due to the large number of observations. The mean 1000-km magnitudes from each Group are plotted versus time in Figure 4.

Table 3. 1000-km magnitudes

```
VisorSat
#    Year      Mag      SD   SD_Mean
4    2020.90   7.30    0.39   0.00
8    2021.12   7.15    0.78   0.01
1    2021.27   6.84    1.05   0.01
9    2021.41   7.52    0.73   0.01
6    2021.52   7.23    1.00   0.01
0    2021.59   7.21    1.03   0.01
Unified        7.21    0.89   0.00
Aggregated     7.21    0.83   0.00
No Scatter             0.81   0.00

Original
#    Year      Mag      SD   SD_Mean
5    2019.79   6.04    0.42   0.00
3    2020.68   5.80    0.40   0.00
2    2021.10   5.84    0.46   0.00
7    2021.48   5.88    0.53   0.01
Unified        5.89    0.46   0.00
Aggregated     5.89    0.45   0.00
No Scatter             0.43   0.00

SD = standard deviation
SD_Mean = SD of the mean
```

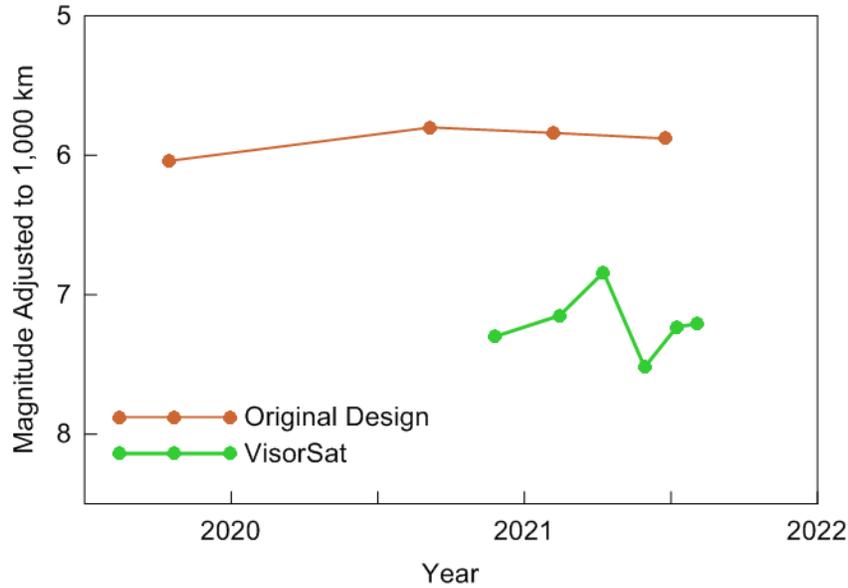

Figure 4. Mean magnitudes of Starlink satellite Groups adjusted to a standard distance of 1,000 km are plotted versus the year of observation.

## 5. Phase function analysis

The standard deviations of Starlink magnitudes around their mean values are quite large as reported in the previous two sections. So, these means are of limited use to astronomers who factor satellite brightness into their observing plans. A more powerful method for quantifying and predicting magnitudes is Phase Function (PF) analysis.

The phase angle provides information about illumination based on the geometry of the satellite, the Sun and the observer (or sensor). An object's phase angle is the arc length between the Sun and the observer as measured at the object. When the phase angle is zero the Sun and observer are aligned as seen from the object, and when it is $180°$ they are in opposite directions.

There is a *standard* PF which corresponds to the fraction of a spherical object illuminated as a function of phase angle. This standard function is brightest at small phase angles when the side of the object facing the observer is fully lit by the Sun and, conversely, it is faintest when the object is back-lit.

An *empirical* PF is derived by least-squares fitting of the observed magnitudes at a standard distance to their phase angles. A quadratic polynomial equation is adequate for fitting Starlink satellite magnitudes.

Figure 5 shows the empirically derived functions for VisorSat and Original satellites along with the standard one. While all three functions are brightest at small phase angles, those for the Starlink satellites depart from the standard function qualitatively. At small and large angles, the satellite PFs substantially exceed the brightness of the standard function. Furthermore, the satellite functions become brighter at large angles where their bodies are back-lit by the Sun as compared to 90$^o$ where they are side-lit. This unexpected behavior indicates strong forward scattering of sunlight. The graph also illustrates that the brightness of VisorSats is more strongly determined by the phase angle than is that for Original satellites.

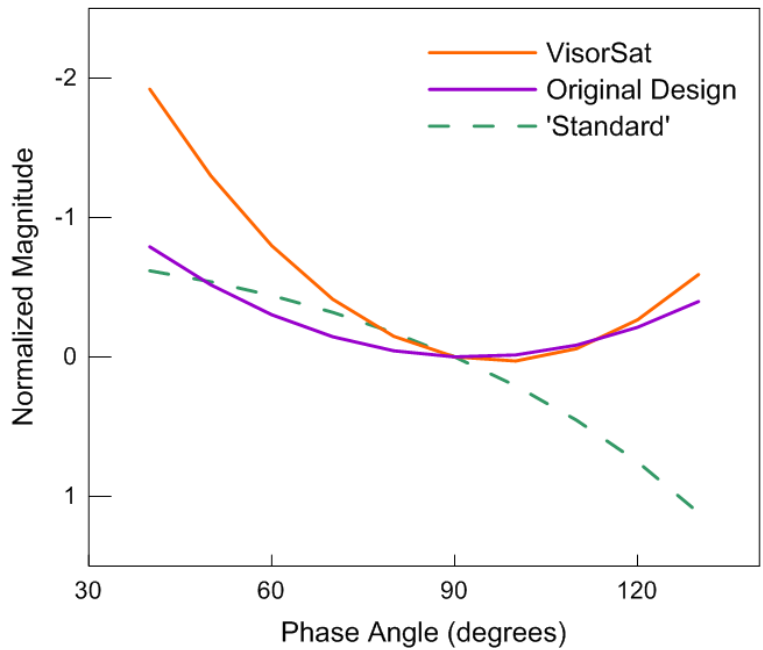

Figure 5. The empirical PFs for VisorSat and Originals are compared to the 'standard'. All three PFs are normalized to magnitude zero at angle 90$^o$.

Analysis by Group number has also revealed that Starlink PFs change with time. Cole (2021) was the first to notice that VisorSats had become brighter in year 2021 when seen at azimuths opposite to the Sun. This same effect can be seen in the PF coefficients which are listed by Group in Table 4. VisorSat Groups 9 and 4 have the largest and smallest coefficients, respectively, for degree 2 of the polynomial equation. These strongest and weakest PFs for VisorSats are plotted in Figure 6. Two possible reasons for the observed variations in PF are adjustment of the attitude (yaw, pitch and roll) of the satellite body and reorientation of its solar array.

Table 4. Phase function coefficients

```
VisorSat      ---- Polynomial Degree ----   - RMS -
#   Year       0        1         2
4   2020.90   5.686   0.04281   -0.0002674    0.38
8   2021.12   2.506   0.10659   -0.0005569    0.54
1   2021.27   2.717   0.08864   -0.0003603    0.49
9   2021.41  -1.980   0.20704   -0.0010546    0.42
6   2021.52   2.382   0.10379   -0.0004993    0.50
0   2021.59   1.221   0.13578   -0.0007116    0.56
Unified       2.058   0.11519   -0.0005907    0.53
Aggregated    2.089   0.11411   -0.0005750    0.48
No Scatter                                    0.44

Original      ---- Polynomial Degree ----   - RMS -
#   Year       0        1         2
5   2019.79   3.770   0.05797   -0.0003497    0.40
3   2020.68   4.598   0.02513   -0.0001086    0.33
2   2021.10   3.617   0.05194   -0.0002727    0.35
7   2021.48   0.655   0.12364   -0.0006850    0.37
Unified       3.646   0.05293   -0.0002857    0.39
Aggregated    3.160   0.06467   -0.0003540    0.36
No Scatter                                    0.33

Coefficients apply to angles in degrees
```

PF analysis reduces the uncertainty of Starlink satellite magnitudes as compared to mean apparent magnitudes and mean 1000-km magnitudes. The first row below the Group results in Table 4, labeled 'Unified', lists the RMS between the function and the observations that results from processing all the observations in one solution. The next line, labeled 'Aggregated', gives the result of combining the RMS values from the individual Group solutions. The Aggregated RMS values are less than the Unified values because the Group solutions vary. Finally, on the line labeled 'No Scatter', the RMS from the Aggregated solution has been adjusted to account for the values of observational scatter listed in Table 1. The No Scatter value of RMS for the VisorSat PF, 0.44, compares to the No Scatter standard deviations of 0.80 and 0.81 for apparent and 1000-km magnitudes, respectively. Likewise, the No Scatter value of RMS for the Original PF, 0.33, compares to the No Scatter standard deviations of 0.49 and 0.43 for apparent and 1000-km magnitudes, respectively.

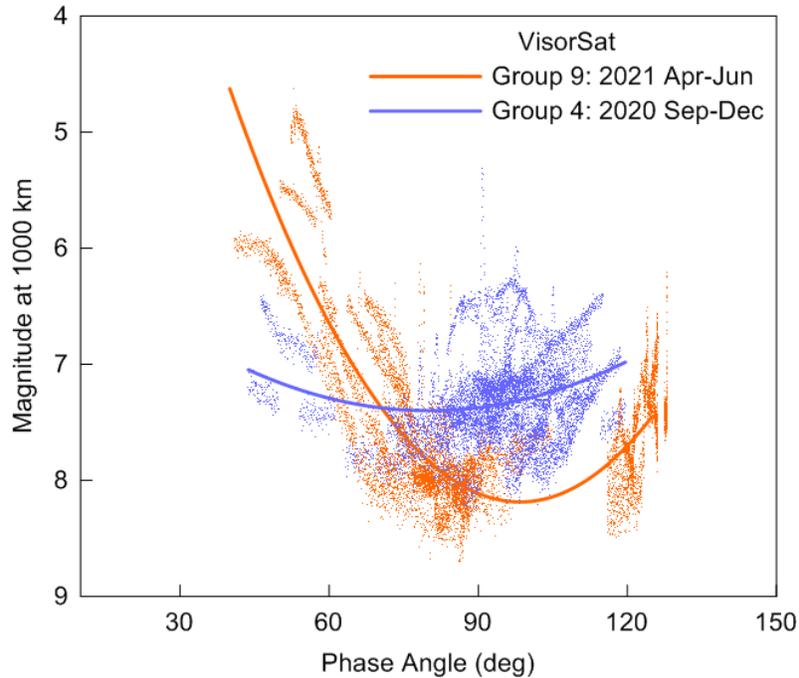

Figure 6. The strongest and weakest PFs for VisorSat. The individual magnitudes and the best fitting quadratic equations are shown.

While a single mean value can characterize the apparent magnitudes and the 1000-km magnitudes there is no corresponding number for the PF. However, the magnitude at phase angle $72^o$ is a good choice for characterization because it corresponds to an object at zenith when the Sun is $18^o$ below the horizon at the end of astronomical twilight. When the PFs derived from all Groups of VisorSat and Original observations are evaluated at $72^o$ the resulting magnitudes are 7.29 and 5.98. When compared to the mean 1000-km magnitudes, these PF values are 0.08 and 0.09 magnitude fainter, respectively.

Finally, there are two unfinished topics from Sections 3 and 4 that are addressed here because they pertain to PFs. First, Group 9 was identified in Section 3 for having the faintest mean apparent magnitude for VisorSats. In this section Group 9 was singled out again because it has the strongest PF. If there is a causal relationship, then a strong PF minimizes interference from bright satellites. Second, Section 4 on 1000-km magnitudes also noted that adjusting apparent magnitudes to a standard distance did not greatly reduce their dispersions as might be expected. The negative values of the degree 2 coefficients of the PFs listed in this section can account for this failure to diminish the standard deviation. When satellites are close to the observer they are overhead and their phase angle is usually in the mid-range. Meanwhile, more distant satellites are nearer to the horizon and their phase angle may be very high or low because the satellites can be nearly in line with the Sun or opposite to the Sun as seen by the observer. Thus, variation of apparent magnitude with distance to the satellite is offset by the PF. Specifically, nearer satellites tend to be in the fainter mid-range of phase angles while only the more distant satellites can be in the brighter extremes of that range.

**6. Starlink Brightness Function**

This section explains the creation and usage of a physical model called the Starlink Brightness Function (SBF). First, a satellite-centered coordinate system (SCCS) that corresponds to the shape of a Starlink spacecraft is defined. The SCCS serves as the frame of reference for the SBF. Then the computation of data needed for the SBF is detailed. This includes the positions of the Sun and the MMT-9 in the SCCS. Next, the evaluation of the parameter coefficients for the SBF is described. Finally, the ability of the SBF to improve magnitude predictions by reducing their uncertainties is demonstrated.

The main physical components of Starlink spacecrafts are a flat-panel shaped body and a large solar array. The body panel is nominally perpendicular to the nadir direction and the solar array is nominally perpendicular to the solar direction when the satellites are at their operational altitude. This geometrical shape and orientation is the motivation for the SCCS. The north pole of the SCCS is in the direction of the satellite zenith and the zero point of azimuth is toward the Sun as illustrated in Figure 7.

The important directions for any physical brightness model are those of the lighting source and of the sensor. In this case they correspond to the Sun and the MMT-9 observatory. The components of the direction vectors are elevation and azimuth, and they are distinct from those in the ground-based coordinate system. The elevations of the Sun and of MMT-9 are measured relative to the equator of the SCCS. The azimuth of the SCCS is defined to be zero in the direction of the Sun. So, the azimuth of MMT-9 is taken to be the difference from that of the solar azimuth.

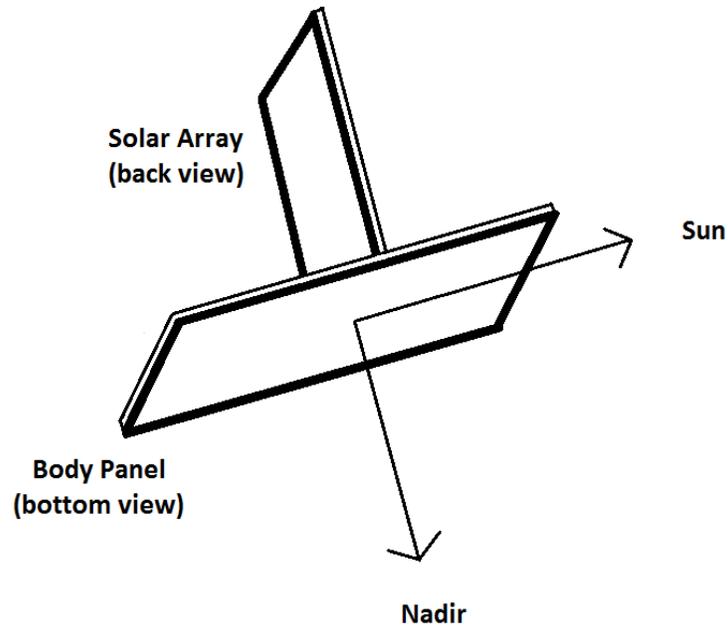

Figure 7. The SCCS frame of reference is illustrated in this schematic diagram of a Starlink spacecraft. The north pole direction is opposite from nadir. The equator is defined by the plane of the body panel. Solar and MMT-9 elevations are generally negative for satellites visible from the ground. The solar direction marks the zero point of azimuth.

The coordinates of the Sun and the MMT-9 site were computed as follows. Two line element sets (TLEs) corresponding to the spacecraft orbit at the time of observation were acquired from Space-Track. The spacecraft Cartesian coordinates in the Earth-Centered Earth-Fixed reference frame were then generated using SPG4 software. The corresponding right ascension (RA) and declination (Dec) values for the north pole of the SCCS were computed and then adjusted for precession to the time of the observation. The Cartesian positions of the MMT-9 site were calculated from its longitude, latitude and height, together with the sidereal time. The precession-adjusted Cartesian coordinates of the satellite were subtracted from those of MMT-9 to determine the RA and Dec of the site which, in turn, were transformed to its azimuth and elevation in the SCCS. The solar RA and Dec and the sidereal time were obtained from the JPL Horizons ephemeris system. The solar coordinates were mapped to azimuth and elevation in the SCCS. Finally, the MMT-9 azimuth difference was calculated by subtracting the solar azimuth.

Coefficients of the three SBF parameters (MMT-9 azimuth, MMT-9 elevation and solar elevation) were determined by fitting them to the 1,000-km magnitudes. Brightness was found to be very sensitive to the MMT-9 azimuth as shown by the quadratic fits in Figure 8. This relationship resembles that of the PFs (see Figures 5 and 6) because the geometries are somewhat similar. In both cases a small value of the independent variable correlates with a large angular separation between the satellite and the Sun as seen from the MMT-9 site, and vice-versa.

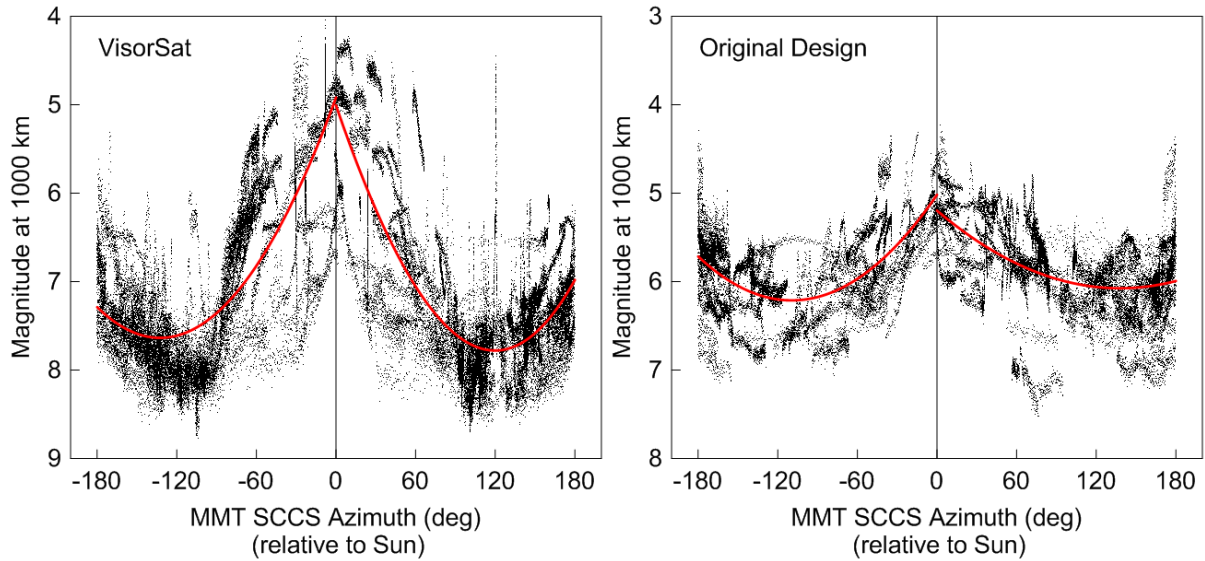

Figure 8. Starlink satellites are brightest when the SCCS azimuths of the Sun and MMT-9 are aligned (0 on the x-axis). The dimming trend reverses beyond $90^{o}$ for both satellite models, but the effect is stronger for VisorSat. Individual magnitudes and the best fitting quadratic equations are plotted.

After azimuth, the next most sensitive parameter is solar elevation which is represented by a linear fitting to the residuals from the azimuth fit. Lastly, MMT-9 elevation was fit linearly to the magnitude residuals from the combined fit of azimuth and solar elevation. The fitting process was iterated 100 times and the resulting coefficients are listed in Table 5 for all VisorSat observations taken together and likewise for all Original observations. The coefficients for each separate Group solution are listed in Appendix B.

The degree 1 coefficient of the fit for solar elevation is a positive number. Since the elevations in the SCCS are generally negative, this indicates decreased magnitude and increased satellite brightness when sunlight impacts the body panel at a more perpendicular angle, as expected. Meanwhile, the degree 1 coefficient for the MMT-9 elevation is negative which implies decreased brightness when the body panel is more perpendicular to the direction of the MMT-9 site. This result was not anticipated but it is probably explained by greater hiding of the large solar array by the body panel as seen from MMT-9 as its elevation in the SCCS approaches $-90^{\circ}$.

Table 5. Coefficients of the SBF parameters

```
VisorSat              ------ Polynomial Degree ------
Parameter                0         1            2
---------             ------   ---------   ----------
Azimuth                4.949     0.04616   -0.0001936
Solar elevation        0.420     0.02580
MMT elevation         -1.095    -0.01931

Original              ------ Polynomial Degree ------
Parameter                0         1            2
---------             ------   ---------   ----------
Azimuth                5.201     0.01604   -0.0000710
Solar elevation        0.910     0.05401
MMT elevation         -0.624    -0.01059

Coefficients apply to angles in degrees
```

Figure 9 illustrates the application of the SBF to the 1000-km magnitudes. The mean VisorSat magnitude at that standard distance is 7.21 and the RMS of the differences of individual observations from that mean is 0.89. When those magnitudes are adjusted with the SBF fitting, the RMS residual reduces to 0.54. For Original satellites the mean and RMS are 5.89 and 0.46, and the RMS reduces to 0.34 after fitting.

Table 6 lists the RMS values for the Unified, Aggregated and No Scatter cases that are described in Section 5. After accounting for Group variations and observational scatter, the RMS residual for VisorSats modeled with the SBF diminishes from 0.54 to 0.44, and that for Originals from 0.34 to 0.24. The corresponding No Scatter RMS values for the PFs are 0.44 for VisorSat and 0.33 for Originals. So, the SBF predicts magnitudes as well as the PFs for VisorSat and improves upon the PFs for Originals.

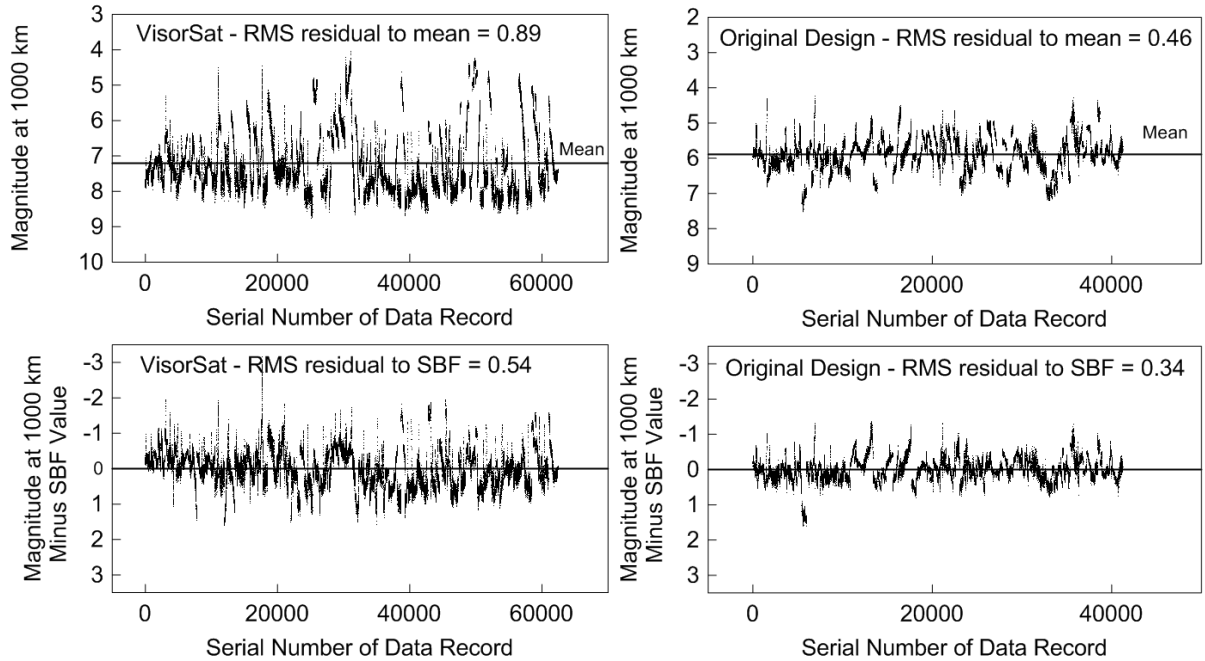

Figure 9. The scatter of magnitudes is reduced when the SBF is applied.

Table 6. SBF values of RMS

| **VisorSat** | RMS |
|---|---|
| Unified | 0.54 |
| Aggregated | 0.48 |
| No Scatter | 0.44 |
| | |
| **Original** | RMS |
| Unified | 0.34 |
| Aggregated | 0.29 |
| No Scatter | 0.24 |

**7. Magnitude prediction**

One reason for characterizing satellite brightness is to inform astronomers who are scheduling observations. The brightest satellites can be avoided by selecting favorable sky areas and optimal times of night. This section explains and illustrates the prediction of Starlink magnitudes.

PF and SBF models used to *characterize* satellite brightness can also *predict* 1000-km magnitudes, and those can be converted to apparent magnitudes by applying the inverse-square law of light. The inputs are the azimuths and elevations of the Sun and of the satellite in the local horizon reference frame as well as the satellite distance.

Examples of predictions for VisorSats based on the PF and SBF models are shown as all-sky maps in Figure 10. Magnitudes at the end of astronomical twilight are represented by colors. Both models indicate that the brightest satellites (magnitude < 5.0, red) are located at mid-range elevations toward the azimuth that is opposite to the Sun. Fainter satellites are represented by adjacent color bands and the Earth shadow region is depicted in black. The corresponding sky maps for Original satellites in Figure 11 demonstrate that those spacecraft exceed the brightness of VisorSats over a much of the sky.

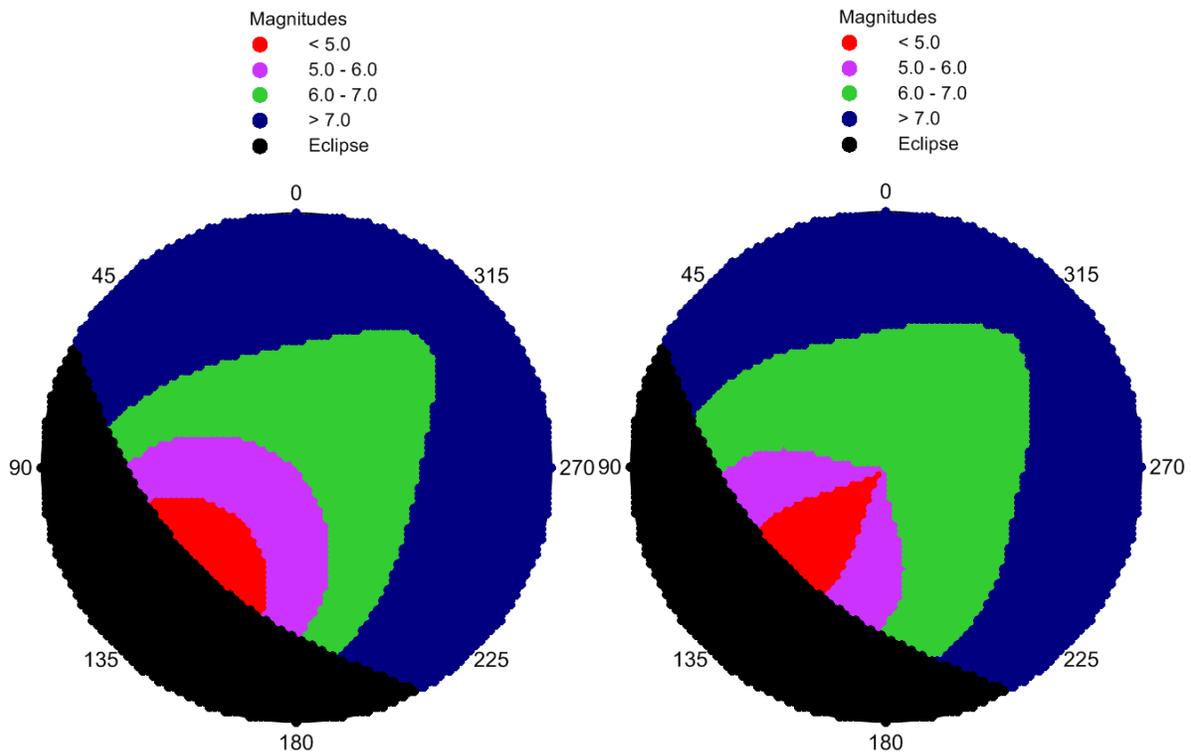

Figure 10. VisorSat brightness over the whole sky as predicted by the PF (left) and the SBF (right) models. The Sun is located 18$^o$ below azimuth 315$^o$.

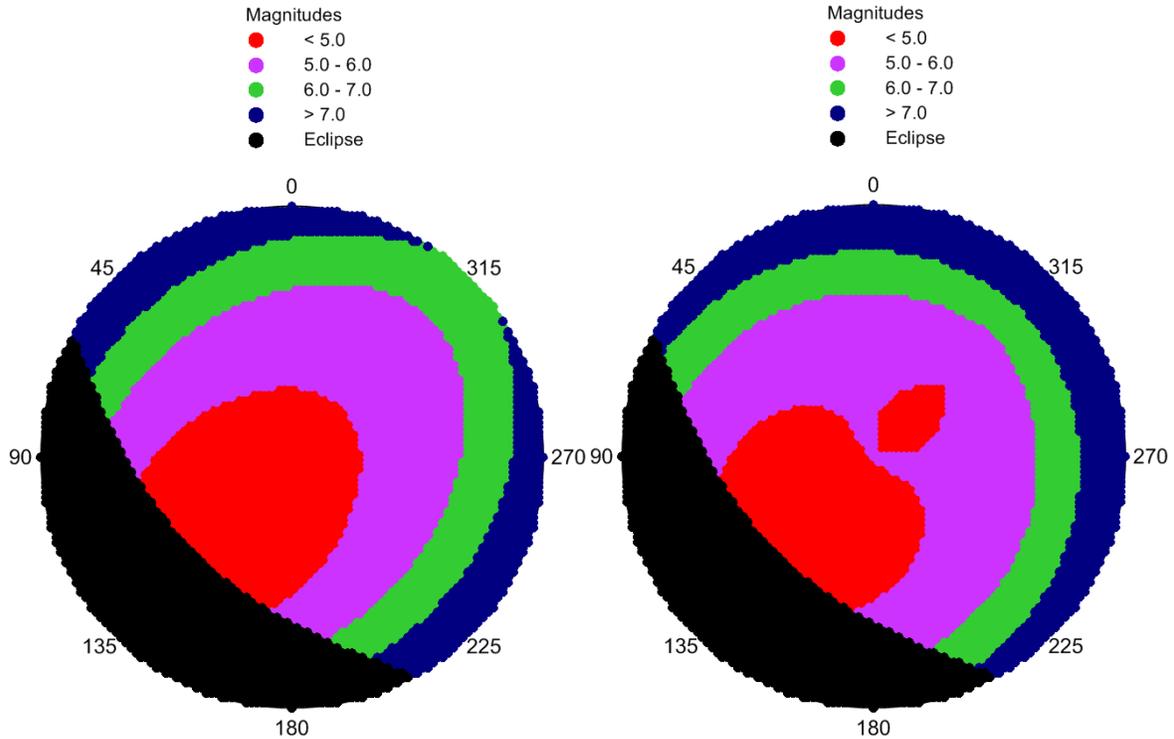

Figure 11. Same as Figure 10 but for Original design satellites. Bright satellites of this type occupy larger sky areas as compared to VisorSats.

PF and SBF magnitude predictions are also useful for generating magnitude statistics and evaluating changes that occur with time of night. Figure 12 plots the percentage of sky occupied by VisorSats brighter than certain threshold magnitudes. Solar elevation drops from $-6^o$ (end of civil twilight) to $-12^o$ (end of nautical twilight) to $-18^o$ (end of astronomical twilight) and finally to $-30^o$. The graph indicates that VisorSats will appear brighter than magnitude 7.0 over more than 80% of the sky above elevation $30^o$ when astronomical twilight ends. Satellites brighter than magnitude 7.0 are eclipsed by the Earth's shadow when the Sun is more than $30^o$ below the horizon.

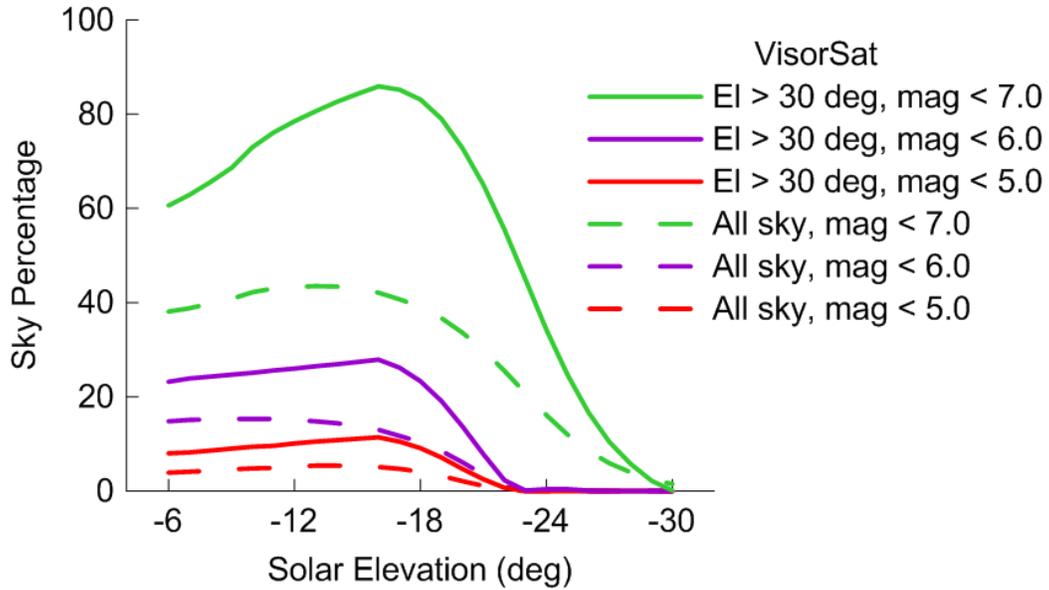

Figure 12. Percentage of sky area with VisorSat magnitudes brighter than the limits indicated. The solid lines represent the sky above 30° elevation and the dashed lines are for the whole sky. Data were generated with the SBF model.

**8. Brightness changes on short time scales**

The intensity of satellite trails on individual astronomical images will change with time and with position. Knowledge of these variations may be useful in developing algorithms to reduce the impact of trails and to reduce loss of data. So, this section describes Starlink magnitude changes that occur on time-scales from seconds to tens of seconds.

The light curve graphs used in this study were inspected by eye for sudden brightness enhancements, called flares. Figure 13 shows one of these symmetrically-shaped surges that usually last a few seconds. Flares are probably due to specular reflection of sunlight from polished surfaces on the satellites.

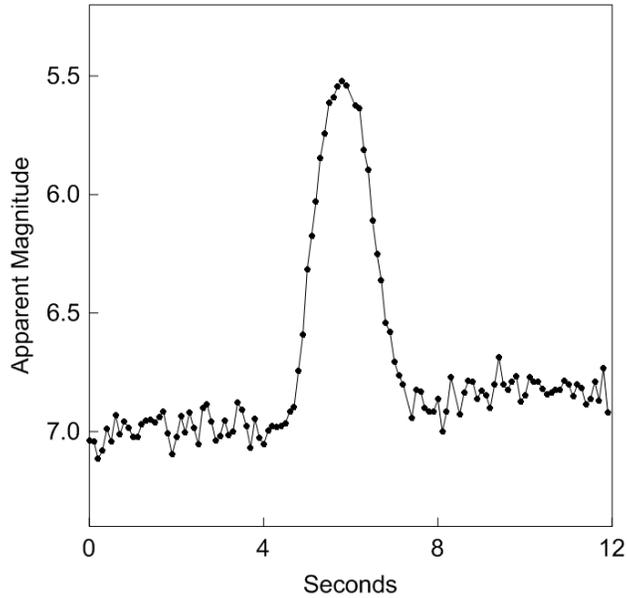

Figure 13. Flaring of Starlink-1538 recorded on 2021 May 2.

The data in Table 7 indicate that VisorSats produce more flares than Originals. The mean intervals between flares that exceed an amplitude of 0.5 in apparent magnitude are 129 seconds for VisorSats and 622 seconds for Originals. The percentage of the elapsed time in all the light curves spent above threshold amplitudes of 0.5, 1.0 and 2.0 magnitudes are also listed in the Table. They range from 0.0% for Original flares exceeding 1.0 magnitude to 2.8% for VisorSat flares of 0.5 magnitude.

Table 7. Flares

```
              Mean Interval    Time Percentage
              (seconds)        -- Magnitude --
                               0.5   1.0   2.0
                               ---   ---   ---
   VisorSat       129          2.8   1.0   0.1
   Original       622          0.4   0.0   0.0
```

More gradual brightness variations were characterized by taking the means of apparent magnitudes in 10 second intervals. This averaging was performed to reduce observational scatter. The mean magnitudes in adjacent time intervals were then differenced and that result was divided by 10. The absolute value of the result was taken to be the rate of change per second. Light curves containing flares were omitted from this analysis.

The same 10-second averages were then used to determine rates of magnitude change per degree of arc distance that the satellite traveled. This result was determined by factoring in the satellite's azimuth and elevation at the mid-point of each interval.

The mean rates of magnitude change are given in Table 8 along with standard deviations and standard deviations of the means. The rates of change per second are somewhat higher for VisorSats than for Originals with means of 0.021 and 0.016, respectively. Likewise, the mean rates of change per degree are 0.038 and 0.033, respectively. The standard deviations are more than half as large as the corresponding means in all cases. This indicates that the rates, despite their rather small size, are highly variable.

Table 8. Gradual brightness variations

|  | Per Second | Per Degree |
|---|---|---|
| **VisorSat** | | |
| Mean rate | 0.021 | 0.038 |
| S.D. | 0.017 | 0.028 |
| S.D. of mean | 0.002 | 0.004 |
| **Original** | | |
| Mean rate | 0.016 | 0.033 |
| S.D. | 0.009 | 0.021 |
| S.D. of mean | 0.001 | 0.003 |

## 9. Failed satellites and other anomalies

This section addresses unusual observations that were encountered during the course of the study. These anomalies can be categorized as failed satellites, particularly bright satellites and satellites in the process of changing altitude.

Failed satellites are identified in the [tables](#) maintained by J. McDowell along with the dates that each one ceased to be operational. Some failed satellites are quickly de-orbited but those considered here are only slowly losing altitude. All three of these cases are Original Starlink satellites. They were encountered when searching for tracks of Original satellites at operational altitudes and they are not included in the analyses reported in previous sections of this paper. The MMT-9 database was searched for all data on these satellites recorded after the date of failure. The mean 1000-km magnitude for each track is plotted in Figure 14 along with a horizontal line indicating the mean magnitude derived for all operational Original satellites in Section 4.

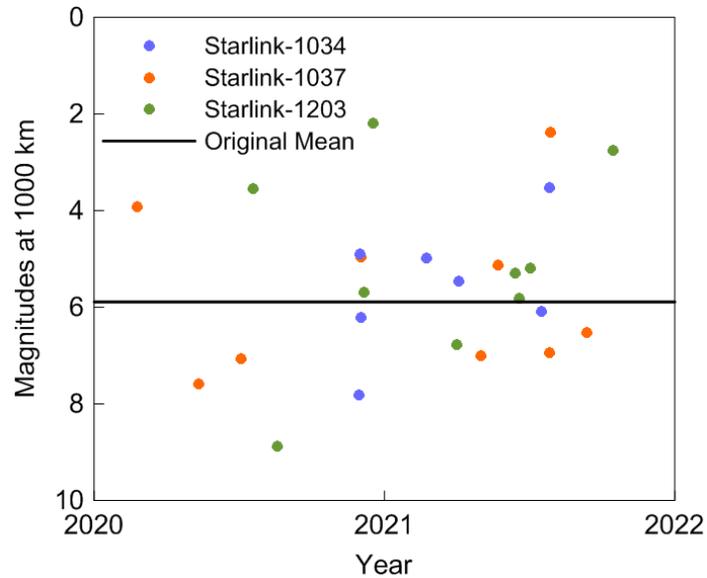

Figure 14. Magnitudes of Original satellites that have failed and the mean magnitude of operational Original satellites.

Table 9 compares the magnitude statistics for the failed satellites to those for all operational Original satellites. The mean 1000-km magnitude for failed satellites is 0.43 brighter than that for the Originals. The standard deviations of the two means differ by only 1.3 sigma, though, which is not statistically significant. On the other hand, the standard deviation for failed satellites, 1.72 magnitudes, is much larger than that for the operational satellites, 0.46. If attitude (yaw, pitch and roll) is uncontrolled for these failed satellites, then the greater variability could be due to their randomized orientations.

Table 9. Magnitudes of failed and operational Original satellites

|  | Failed | Operational |
|---|---|---|
| Mean | 5.46 | 5.89 |
| St. Dev. | 1.72 | 0.46 |
| SD Mean | 0.34 | 0.00 |

The brightness of Starlink satellites is highly variable even within the categories of VisorSat and Original. This magnitude dispersion has led to speculation that some individual satellites are intrinsically brighter than others. Lawler et al. (2021) and Boley et al. (2021) have raised the intriguing question of whether the Sun shades on VisorSats all deployed successfully.

The question of intrinsic variability was investigated by identifying VisorSats whose 1000-km magnitudes were brighter than the mean for that type and whose residuals after SBF fitting were also brighter. This search was conducted using graphs like those in Figure 9, where deviations greater than one magnitude were sought.

The MMT-9 database was searched for all observations of the 10 candidate 'bright' satellites listed in Table 10. Statistics of the mean 1000-km magnitudes for each track of data recorded at the operational altitude are listed in the Table. The value of *sigma* is equal to the difference (*delta* in the table) between the mean 1000-km magnitude for a candidate satellite and the mean for all VisorSats, divided by the standard deviation of the mean (*SDM* in the table). Only one satellite, Starlink-1673, has a statistically significant difference but, unfortunately, there are only 3 tracks of data for that object. So, observational evidence for intrinsically bright VisorSats is not robust.

Table 10. Candidate 'bright' VisorSats

| Starlink Number | Tracks | 1000-km magnitudes Mean | Delta | SDM | Sigma |
|---|---|---|---|---|---|
| **1673** | **3** | **6.25** | **−0.96** | **0.17** | **−5.6** |
| 1550 | 5 | 7.23 | 0.03 | 0.72 | 0.0 |
| 1561 | 6 | 6.78 | −0.43 | 0.24 | −1.8 |
| 1586 | 7 | 7.34 | 0.13 | 0.20 | 0.7 |
| 1721 | 3 | 7.47 | 0.26 | 0.32 | 0.8 |
| 1707 | 3 | 5.84 | −1.37 | 0.94 | −1.5 |
| 1800 | 9 | 7.20 | 0.00 | 0.18 | 0.0 |
| 1872 | 10 | 6.89 | −0.31 | 0.28 | −1.1 |
| 2101 | 4 | 6.61 | −0.60 | 0.49 | −1.2 |
| 2113 | 6 | 7.46 | 0.25 | 0.24 | 1.1 |

This section ends with a brief and anecdotal look at satellites whose orbital altitudes are changing. Mallama (2021) noted that VisorSats are about 0.5 magnitude brighter than normal for a short time after they complete their orbit-raising and attain the operational altitude. The analysis reported in this paper purposely excluded observations of satellites that had only been at the operational altitude for a brief time, as well as those from satellites which had not yet reached that height. These exclusions were based on altitude [plots](plots) maintained by J. McDowell. Those same graphs also reveal periods of time when the altitudes of certain satellites already at their operational height were changing. Two satellites were very bright during such times. The mean 1000-km magnitude for the track of Starlink-1572 recorded on 2021 May 19 was 4.08 and that for Starlink-1591 on 2020 December 5 was 2.93. Starlink-1591 reached apparent magnitude 1.85 which is the brightest value reported in this study.

**10. Limitations of this study**

The general applicability and the accuracy of the results from this study are limited by several factors. First is that the satellite bodies are only *nominally* aligned with the SCCS frame of reference, while their *actual* attitude (yaw, pitch and roll) at the time of each observation is unknown. So, the SCCS angles may not be exactly aligned with the SBF parameters corresponding to the physical spacecraft.

Furthermore, the physical and optical modeling of the spacecraft is only approximate. There is no modeling for the visor or any of the other small spacecraft attributes. Likewise, the light scattering properties of the satellites surfaces are not specified.

Another limitation is that radio communication VisorSat spacecraft are not analyzed separately from the newer laser communication type. The brightness characteristics of these two models might differ due to the distinct hardware features.

A further limitation is that the data are all from one location. If spacecraft attitude depends upon geographic latitude and longitude then the magnitudes acquired by MMT-9 in Russia may not be representative of other regions.

Finally, only Starlink satellites orbiting at 550 km altitude and 53$^o$ inclination are addressed. While that includes almost all spacecraft launched to date, many satellites will be placed in other orbits during the coming years. Those may have different brightness characteristics.

## 11. Discussion

This section places the analysis reported here in the context of other photometric research involving Starlink satellites. Several talks given at the *Dark and Quiet Skies For Science and Society II* conference provided current information on this rapidly changing topic. Hainaut et al. (2021) analyzed the increase in artificial sky brightness from satellite constellations. Tregloan-Reed (2021) summarized the development of a global observing network. This association is centered in Chile but also receives photometric data from observations from Spain, Vietnam and South Korea.

Two papers at the conference provided mean magnitudes based upon thousands of Starlink observations. In the first, Otarola et al. (2021) reported on data obtained with the Zwicky Transit Facility in the Sloan g-, r- and i-bands. They find two distinct brightness groupings which they identify with Original and VisorSat spacecraft. Their mean Sloan magnitudes were converted to V-band values for comparison with MMT-9 results using the equations of Smith et al. (2002) and then scaled to 1000 km. (MMT-9 magnitudes are within about 0.1 magnitude of V as demonstrated in Appendix A.) The resulting VisorSat and Original V-band means are 7.40 and 6.35, respectively. These results are fainter by 0.19 and 0.46 magnitudes than the corresponding means reported in this paper.

In the other conference paper, Scott et al. (2021) reported on a space-based photometric study of satellite constellations. The mean of their 'open filter' Starlink magnitudes at the 1000 km distance is 6.9. They do not distinguish between the two types of Starlink satellites. Their mean is 0.35 magnitude fainter than the average of the means of VisorSats and Originals in this paper.

While the Starlink magnitudes reported at the *Dark and Quiet Skies* conference agree with those from this paper within 0.5 magnitude, Krantz et al. (2021) report considerably fainter results. They give 'as-observed, only corrected for airmass extinction' magnitudes which are like the apparent magnitudes listed in this paper. Their means of 8.0 for VisorSats and 7.0 for Originals are 1.6 and 2.0 magnitudes dimmer than the corresponding means in this study. Furthermore, the analysis in Section 7 of this paper indicates that VisorSats will appear brighter than magnitude 7.0 over more than 80% of the sky above 30$^o$ elevation when astronomical twilight ends. So, there is a substantial difference between the results of the two studies.

Krantz et al. attribute their faint values to 'an increased number of measurements low in elevation' and point out that 'more satellites are visible low in elevation'. These two points can explain the dim magnitudes resulting from their all-sky observing strategy. Their paper does not report an average satellite elevation but the geometry of a hemisphere gives some indication. Half the surface area of a hemisphere is above 30$^o$ and half is below, which suggests a value for the mean elevation of an all-sky survey. By contrast, the mean satellite elevations for the photometry reported in this paper are considerably higher at 53$^o$ for VisorSat and 56$^o$ for Originals.

**12. Conclusions**

More than 100,000 Starlink magnitudes recorded by the MMT-9 robotic observatory were analyzed. These visible light values represent brightness from the blue part of the spectrum through the red, and they are within about 0.1 magnitude of the V-band.

The mean apparent magnitude for VisorSats is 6.43 and that for Original-design models is 5.05. The standard deviations to the means are 0.86 and 0.52 magnitude, respectively. The formal uncertainties of the mean values are less than 0.01 magnitude.

The mean of magnitudes adjusted to a range of 1,000 km is 7.21 for VisorSats and 5.89 for Originals. The difference of 1.32 magnitudes implies that VisorSats average 30% as bright as the Originals. The standard deviations are 0.89 and 0.46 magnitude, and the formal uncertainties are less than 0.01 magnitude. There is no large and overall reduction in the standard deviations compared to those of apparent magnitudes.

The phase functions (PF) for both models of satellite indicate strong forward scattering of sunlight. They are also time-dependent on scales of months and years. The RMS residuals to the time-dependent phase functions are 0.48 for VisorSat and 0.36 for Original. When observational scatter is taken into account these RMS values reduce to 0.44 and 0.33.

A time-dependent Starlink Brightness Function (SBF) tailored to the shape and orientation of the satellites fits the magnitudes for Original design satellites with an RMS residual of 0.24 after observational scatter is removed. This is better than the RMS for the PF. However, for VisorSats, the RMS values from the SBF and from the PF are the same at 0.44 magnitude.

Brightness flares that exceed 0.5 magnitude occur at an average frequency of once per 129 seconds for VisorSats and those satellites are in that elevated state 2.8% of the time. Such flares for Original design satellites are less frequent, occurring once per 622 seconds and lasting only 0.4% of the time.

The rates of gradual brightness change (not including flares) and their standard deviations over the course of a pass average 0.038 +/- 0.028 magnitude per arc-degree of motion for VisorSats and 0.033 +/- 0.021 for Originals. The corresponding values per second of time are 0.021 +/- 0.017 magnitude and 0.016 +/- 0.009, respectively. The standard deviations of the means are all in the milli-magnitude range.

Several satellites that have failed in orbit exhibit very wide ranges of brightness. No strong evidence was found for individual satellites that are intrinsically brighter than average. Satellites whose orbital altitudes are changing can be very bright.

The mean 1000-km magnitudes found in this study are consistent with those derived from the results presented by Otarola et al. (2021) and with those listed by Scott et al. (2021). However, Krantz et al. (2021) report much fainter mean apparent magnitudes than those in this paper. That difference is attributed to the lower elevations of the satellites sampled by Krantz et al.

Magnitude prediction is discussed in the context of planning astronomical observations. Sky maps with satellite brightness as a function of azimuth and elevation are illustrated. Graphs showing satellite magnitudes as a function of the time of night are also shown.

Acknowledgment

D.R. Skillman verified spherical geometry transformations for the satellite-centered coordinate system.

## Appendix A. MMT-9 and V-band magnitude transformation

Karpov (private communication) indicated that '[MMT-9] Instrumental ("Clear") magnitudes are in the system which is roughly Johnson V + 0.15 * (B-V)'. So,

$$V = M_C - 0.15 * (B-V)$$

Equation A-1

where $M_C$ is the MMT-9 clear magnitude. For the solar color index of 0.63,

$$V = M_C - 0.09.$$

Equation A-2

Thus, the MMT-9 clear magnitude is within approximately 0.1 magnitude of the V-band for grey bodies reflecting sunlight.

## Appendix B. Starlink Brightness Function coefficients

VisorSat – time ordered

```
                   - Polynomial Coefficient -
Parameter            0       1        2
---------          ------  -------  ----------

Group 4
Azimuth             6.324   0.02668 -0.0001343
Solar elevation     0.587   0.03525
MMT elevation      -0.229  -0.00369
RMS                 0.260

Group 8
Azimuth             5.608   0.03004 -0.0001171
Solar elevation     0.630   0.04017
MMT elevation      -1.843  -0.03132
RMS                 0.479

Group 1
Azimuth             4.949   0.04288 -0.0001726
Solar elevation     0.992   0.05986
MMT elevation      -0.990  -0.01771
RMS                 0.508
```

```
Group 9
Azimuth                    5.468   0.03969 -0.0001604
Solar elevation            0.101   0.00672
MMT elevation             -1.015  -0.01846
RMS                        0.545

Group 6
Azimuth                    4.381   0.05915 -0.0002578
Solar elevation           -0.209  -0.01179
MMT elevation             -0.622  -0.01160
RMS                        0.570

Group 0
Azimuth                    4.499   0.05374 -0.0002218
Solar elevation            0.062   0.00388
MMT elevation             -1.699  -0.03083
RMS                        0.458
```

Original – time ordered

```
                        - Polynomial Coefficient -
Parameter                 0       1         2
---------               ------  ------- ----------

Group 5
Azimuth                   5.031   0.02469 -0.0001138
Solar elevation           0.558   0.03061
MMT elevation             0.114   0.00175
RMS                       0.340

Group 3
Azimuth                   5.529   0.00537 -0.0000194
Solar elevation           0.912   0.05790
MMT elevation            -0.652  -0.01244
RMS                       0.320

Group 2
Azimuth                   5.292   0.01626 -0.0000801
Solar elevation           1.337   0.08117
MMT elevation            -0.179  -0.00283
RMS                       0.229

Group 7
Azimuth                   4.501   0.03131 -0.0001410
Solar elevation           1.460   0.08598
MMT elevation            -0.993  -0.01813
RMS                       0.246

Coefficients apply to angles in degrees
```